\documentclass[doublecol]{epl2} 

\title{Growth of a dynamical correlation length \\in an aging superspin glass}
\shorttitle{Extraction of a dynamical correlation length in a superspin glass} 

\author{E. Wandersman\inst{1} \and V. Dupuis\inst{1} \and E. Dubois\inst{1} \and R. Perzynski\inst{1} \and S. Nakamae\inst{2}  \and E.Vincent\inst{2}  }
\shortauthor{E. Wandersman \etal}

\institute{                    
  \inst{1} Laboratoire des Liquides Ioniques et Interfaces Charg\'ees \\ UMR 7612 CNRS - Universit\'e Pierre et Marie Curie - ESCPI \\ 4 place Jussieu - Bo\^ite 51 - 75252 Paris Cedex 05 - France\\
  \inst{2} Service de Physique de l'Etat Condens\'e (CNRS URA 2464)\\ DSM/IRAMIS/SPEC - CEA Saclay, 91191 Gif sur Yvette Cedex, France
}

\abstract{
We report on zero field cooled magnetization relaxation experiments on a concentrated frozen ferrofluid exhibiting a low temperature superspin glass transition. With a method initially developed for spin glasses, we investigate the field dependence of the relaxations that take place after different aging times. We extract the typical number of correlated spins involved in the aging dynamics. This brings important insights into the dynamical correlation length and its time growth. Our results, consistent with expressions obtained for spin glasses, extend the generality of these behaviours to 
the class of superspin glasses. Since the typical flipping time is much larger for superspins than for atomic spins, our experiments probe a time regime much closer to that of numerical simulations.
}

\pacs{75.50.Lk}{\textit{Spin glasses and other random magnets}}
\pacs{75.50.Tt}{\textit{Fine-particle systems; nanocrystalline materials}}
\pacs{75.40.Gb}{\textit{Dynamic properties}}

\begin{document}

\maketitle

\section{Introduction}
The microscopic mechanisms which are at the origin of the dramatic slowing down observed at the glass transition in a variety of complex systems, such as structural glasses or frustrated and random magnets, remain poorly understood \cite{KobLivre}.  Nevertheless, recent developments \cite{BerthierI} relate these slow dynamics to the cooperative nature of the dynamical processes: for a particle (or a spin) to move, a part of its neighborhood must be engaged in this motion. The particle dynamics in glasses is then heterogeneous both in time and space, leading to anomalous dynamical fluctuations. The intensity of dynamical fluctuations gives an access to the size of these dynamically correlated domains.\\
Several theoretical and experimental works have been carried out on the dynamical fluctuations in structural glass formers and spin glasses \cite{CugliandoloArXiv04}. Relevant physical quantities are four-point correlators \cite{Berta,Bertb,RiegerBook,Komori:JPSJ99,Marinari:PRL96,Marinari:JPhysA00} or their volume integrals $\chi_4$, the - so called - dynamical susceptibility. Experimental results on structural glasses, although being scarce, begin to draw a universal picture \cite{TarjusArxiv07,ScienceElMasri}, independent of the details of a considered system: as the glass transition is approached, the number of correlated particles increases but remains modest, around 5 to 10 particles.\\
Another interesting question concerns the connexion between structural glasses (with annealed, i.e. self induced, disorder) and spin glasses (with quenched disorder). In a recent paper \cite{Coniglio07}, Coniglio and coworkers show that in the case of quenched disorder, the dynamical susceptibility grows with lag time and reaches a plateau, whose height increases with decreasing T. The deduced dynamical correlation length is diverging at low temperature, \emph{in the same way as} the static non-linear susceptibility. This behaviour is characteristic of a static phase transition. On the contrary, for annealed disordered systems, the authors of \cite{Coniglio07} obtain a peaked dynamical susceptibility, that equals zero at large times. This fact is related to the transient nature of dynamical heterogeneities in this class of systems.\\
Lastly, being out of thermodynamical equilibrium, the dynamics of glassy systems evolves throughout time. The dynamics becomes slower and slower with elapsed time: the system ages. A challenging question in this field is to relate aging to the growth of a correlation length $\xi(t_w)$. However, the time invariance breaking complicates the task of experimentalist to extract this time-dependent dynamical correlation length and therefore, no data is available - to our knowledge - in structural glass formers. In the case of spin glasses, alternative experimental methods have been employed to extract a correlation length, which indeed grows as the system ages \cite{JohPRL99,DupuisPRL04}. From the theoretical point of view, several developments have been achieved to study this growing lenghtscale associated to aging\cite{Jaubert07,RiegerBook}. Recently, Jaubert and coworkers computed local four-point correlators in a 3D Edwards-Anderson spin glass model, at $T\sim0.7T_g$ \cite{Jaubert07}. The deduced correlation length, of the order of a few lattice spacings, is increasing with the waiting time $t_w$ (and diverge in the asymptotic limit), following a power law behaviour $\xi(t_w)\sim t_{w}^{a}$, with an exponent $a\sim0.1$, in agreement with previous numerical works \cite{Berta,Bertb,RiegerBook,Komori:JPSJ99,Marinari:PRL96,Marinari:JPhysA00}.  
\par In the present work, we deal with a peculiar system, namely strongly interacting magnetic nanoparticle dispersions. Each nanocolloid bears a permanent magnetic moment and these systems can thus undergo two distinct kinds of glass transition, considering either structural or magnetic degrees of freedom. The first type of glass transition takes place at \emph{ambient temperature} in concentrated ferrofluids, and is governed by the volume fraction. At high concentrations, a disordered colloidal solid of nanoparticles is observed (annealed disorder). The dynamical properties of this colloidal glass (and more precisely, their heterogeneous nature) are reported~\cite{Wandersman07,GuillaumeBiref}.\\
The second type of glass transition, on which we will focus here, occurs at \emph{low temperature}, when the carrier liquid of a repulsive colloidal dispersion is frozen  in zero magnetic field (quenched disorder). Magnetic nanoparticles are thus blocked  in a solid matrix, randomly distributed and orientated, with each particle being magnetically uniaxial \cite{GazeauRFM}. The only remaining degrees of freedom are the magnetic moment orientations, guided by (i) the individual magnetic anisotropy and (ii) dipolar interaction among magnetic moments, which becomes important at low temperature and sufficiently high volume fraction. Due to the disorder of the nanoparticle positions and magnetic axes orientations, the sign and the strength of the dipolar interactions are random. At low temperature, a transition towards a frustrated state of randomly interacting giant spins is observed, with slow dynamics and aging. These observed similarities with spin glasses \cite{Mydosh:SG, Sitges, UppsalaReview, Jonsson:PRL98, Mamiya:PRL98,Jonsson:JMMM01, Sun:PRL03, Parker:JAP05,ParkerPRB08} are at the origin of the term `superspin glasses' which is now used to describe such materials \cite{Jonsson:ACP04}. \\
Our present aim concerns this superspin glass transition and more precisely the extraction of a growing number of correlated superspins as the system ages. We also compare our results with experimental as well as numerical results obtained for spin glasses.
 The paper is organized as follows. In a first section, we present the analytical method, already employed in spin glasses, to extract the typical number of correlated superspins. We then describe the sample and the experimental procedure used in this study. The results are shown in a following section and are finally discussed.

\section{Analytical and experimental methods}

For spin glasses an indirect method relying on the use of magnetic field change experiments has been successfully used to estimate the number of spins participating in the slow cooperative aging dynamics \cite{Vincent:PRB95, DupuisPRL04,JohPRL99,VincentBook}. Here, we measure the relaxation of the Zero Field Cooled Magnetization for various field amplitudes (ZFCM method). The main idea behind this method is as follows. After a quench in $H=0$ down to a temperature $T_m$ in the spin glass phase (the first stage of a typical ZFCM relaxation experiment) spin glass correlations slowly develop among spins. During the waiting time $t_w$ (counted as soon as the temperature $T_m$ is stable) the system is exploring its metastable configurations in search for an equilibrium state. The typical free energy barrier $B(t_w)$ that can be overcome after $t_w$ involves the cooperative flip of  a number of correlated spins $N_s(t_w)$ which is increasing with the age $t_w$. If a magnetic field $H$ is applied at time $t_w$ (second stage of a ZFCM relaxation experiment, defines $t=0$) the Zeeman energy, $E_Z(H)$, which results from the coupling of the $N_s(t_w)$ spins to the field, is expected to depend on $N_s$ and reduces the typical barrier $B(t_w)$ to $B(t_w)-E_Z(H)$. For vanishing $E_Z(H)$ (i.e. small $H$) one can expect (e.g. in the framework of Bouchaud trap models \cite{BouchaudTrap}) a maximum in the relaxation rate $S=dM/dlog\ t$ at a characteristic time $t \sim t_w$. In the presence of non negligible $E_Z(H)$ one expects a shift of the position of this maximum to shorter times $t$. This shift can be described as a reduction of the true age of the system $t_w$ to an effective age $t_w^{eff}(H)$ :

\begin{equation}
t_w^{eff}(H)=\tau_0 exp\left[\frac{B(t_w)-E_Z(H,N_s(t_w))}{k_B T}\right]
\end{equation}

\noindent or equivalently with $t_w=\tau_0 e^{\frac{B(t_w)}{k_{B}T}}$

\begin{equation}
t_w^{eff}(H)=t_w exp\left[-\frac{E_Z(H,N_s(t_w))}{k_B T}\right]
\end{equation}

\noindent where $\tau_0$ is a microscopic flipping time, of the order of 10$^{-12}$ s  in the case of spin glasses. For magnetic nanoparticles, it is usually taken of the order of 10$^{-9}$ s, but at low temperature, the effect of the anisotropy barrier may be to increase it by several orders of magnitude.\\
According to this scenario, ZFCM relaxation measurements performed at $T_m$ in the glassy phase with probing fields, $H$, of increasing intensities give access to the Zeeman energy $E_Z(H,N_s(t_w))$. The Zeeman energy is
\begin{equation} 
E_{Z}\left( H,N_s(t_w) \right)=M\left(H,N_s(t_w)\right).H
\end{equation}
\noindent $M$ being the magnetization of the set of $N_s$ (super-)spins.\\
 The explicit dependence of $M$ on $N_s$ is not completely obvious for a disordered system. For a small number of Ising spins $N_s$ in a random configuration, the magnetization is proportional to the typical fluctuation $N_s^{1/2}$, and is independent of the field: 
\begin{equation} 
E_Z= N_s^{1/2}\mu H 
\end{equation}
\noindent where $\mu$ stands for the magnetic moment of 1 (super-)spin in the compound. \\
On the other hand, at the macroscopic scale, the magnetization is an extensive quantity, proportional to the number of (super-)spins, and (to first order) proportional to the field via the susceptibility $\chi_{FC}$ of 1 (super-)spin: 
\begin{equation}
E_Z= N_{s} \chi_{FC} H^2
\end{equation}
This quadratic dependence has been found to better fit to the results for Heisenberg spins [7]. We confront below our data to both (quadratic or linear) scenarii.
\par From the $H$ dependence of $E_Z$ one can thus extract the typical number of correlated spins participating in a spin glass dynamics. $N_s$ can then be used to estimate a dynamical correlation length, with the simplest approximation being $N_s \sim \xi^3$, as assumed in\cite{JohPRL99,DupuisPRL04}. However, in numerical simulations \cite{Berta,Bertb,RiegerBook,Komori:JPSJ99,Marinari:PRL96,Marinari:JPhysA00}, 4-point correlation functions can be directly determined. It is found that the equilibrium correlation function involves a power law prefactor $1/r^\alpha$, suggesting that the spins which flip coherently constitute a "backbone" of fractal dimension $d-\alpha$. The calculated values of $\alpha$ are $\sim0.5$ for Ising spins \cite{Berta,Marinari:PRL96} and $\sim1$ for Heisenberg spins \cite{Bertb}, hence yielding $N_s \sim \xi^{2-2.5}$. Clearly, in the present experiments without the knowledge of the effective dimension of the spin backbone, only $N_s$ can be determined. The precise values that we can infer for the dynamical correlation length $\xi$ depends on the backbone dimension.\\
\par
The sample used here is a chemically synthesized concentrated ionic ferrofluid \cite{Massart:IEEE81}, based on maghemite $\gamma-Fe_2O_3$ nanoparticles dispersed in glycerin. The  nanoparticle surface is negatively charged, the dispersion being thus electrostatically stabilized. Each nanoparticle bears a permanent magnetic moment $\mu \sim 10^4 \mu_B$. The particles are slightly polydisperse in size with a diameter following a lognormal law with median diameter $d_0=8.6\ nm$ and polydispersity $\sigma = 0.23$ (estimated from a size weighted Langevin fit of M(H) measurements at room temperature). The nanoparticle volume fraction $\Phi$ is determined to 15\%. The physico-chemical control of the dispersion (ionic strength, osmotic pressure, etc.)  allows  a fine tuning of the electrostatic repulsions (in regard to the dipolar interaction), leading to well dispersed nanoparticles, without the presence of large aggregates or chains.This sample is similar to a previously studied water based frozen ferrofluid \cite{Parker:JAP05} but with a lower volume fraction. 

The measurements reported here are performed with a commerical $Cryogenic^{Ltd}$ S600 and QuantumDesign MPMS SQUID magnetometers. Prior to the measurements, the sample (mass $\approx$ 2 mg) is inserted in a glass capillary (diameter = 1 mm). The sample is introduced in the SQUID and cooled without the presence of any magnetic field, assuring a disordered orientational distribution of the nanoparticle anistropy axes. Note also, that all the results presented here are done at temperature below the fusion point of glycerin, assuring that the nanoparticle orientations are blocked in a rigid matrix.\\
In order to first characterize our sample and evidence the low temperature superspin glass transition of the concentrated frozen ferrofluid, Field Cooled (FC), Zero Field Cooled (ZFC) dc susceptibilities as well as real and imaginary parts of ac susceptibility at frequencies in the range 0.08 - 800 Hz are measured as a function of temperature. Then, to access the dynamical correlation length, we perform ZFCM relaxation experiments at a temperature $T_m=0.7\ T_g$ for probing fields of increasing intensities in the  range [0 - 8.5 Oe]. Note that in references \cite{Vincent:PRB95, DupuisPRL04,JohPRL99}, a TRM protocol was used, which is the mirror procedure of ZFCM (the magnetic field is applied during $t_w$, then turned off; the magnetization relaxation is recorded).\\
But, as noted in  \cite{SasakiPRB05}, the field-cooled state of frozen superparamagnetic nanoparticles is strongly out of equilibrium, inducing possible aging-like effects even for non-interacting nanoparticles. We have therefore preferred to use a ZFCM procedure, in which the sample is cooled in zero field. This choice does not affect the argument on the Zeeman energy, presented above.    

\section{Results}

\subsection{dc and ac susceptibilities}

Figure 1 shows the Field Cooled (FC) and Zero Field Cooled (ZFC) susceptibilities $\chi_{FC}$ and $\chi_{ZFC}$ as well as the real part, $\chi'$, and imaginary part, $\chi''$, of the ac susceptibility at various  $\omega$, in the 0.08 - 800 Hz range as a function of temperature. In each case, a $1\ Oe$ probing field is used. While at high temperature $\chi_{FC}$, $\chi_{ZFC}$ and $\chi'$ exhibit a typical paramagnetic-like Curie Weiss behaviour, $\chi''$ being essentially zero, a transition towards a disordered and frustrated state is observed below $T_g=67\ K$ at which $\chi_{ZFC}$ displays a cusp and below which $\chi_{FC}$ becomes nearly temperature independent. Looking at the ac susceptibility we observed that both $\chi'$ and $\chi''$ exhibit peaks (at slightly different temperatures as expected from Kramers-Kr\"onig relations) which shift towards higher temperatures with increasing pulsation $\omega$.
\begin{figure}[h!]
\includegraphics[width=8cm]{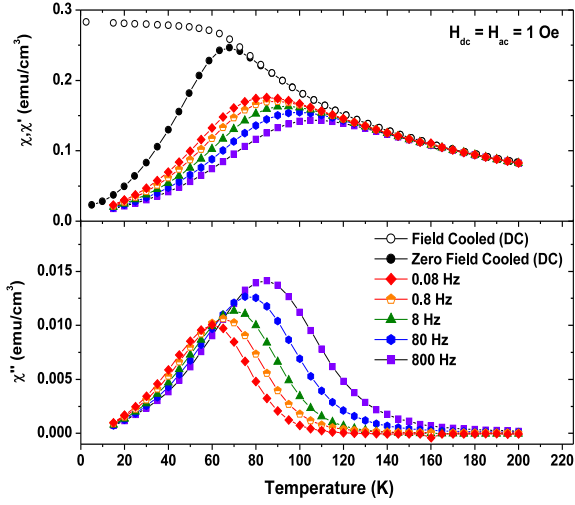}
\caption{\label{fig1} Field Cooled, Zero Field Cooled and real and imaginary parts of resp. dc and ac susceptibilities as a function of temperature.}
\end{figure}
 In order to discriminate between simple superparamagnetic blocking and a true superspin glass transition \cite{Tholence:Physica84} we analyze the $\omega$-shift of the position of $\chi'$ peak using both an Arrhenius law, $\frac{1}{\omega}=\tau_0 e^{\frac{E_a}{k_{B}T_g(\omega)}}$, and a critical law, $\frac{1}{\omega}=\tau_{0}(\frac{T_g(\omega)}{T_g}-1)^{-z\nu}$.\\
 We find here that the Ahrrenius analysis leads to unphysical parameters ($\tau_0\sim10^{-17}s$) whereas the critical analysis can reproduce our observations with reasonable values : $z\nu\approx 7$, $\tau_0\approx 5. 10^{-6}$ s. The rather long time scale obtained for $\tau_0$ is likely to be related to the individual slowing down of the superspin flip, due to the individual anisotropy barriers of the nanoparticles. Indeed, for superspins the more pertinent individual time is $\tau_0\sim\bar{\tau_0} e^{\frac{E_a}{k_{B}T}}$, where $\bar{\tau_0}$ is a typical microscopic attempt time. Referring to data from \cite{GazeauRFM}, obtained with similar nanoparticles, the anisotropy energy of an individual nanoparticle is $E_a/k_B$=2x300 K, and therefore their flipping time at the temperature $T_g$=67 K and with an attempt time $\bar{\tau_0}\sim10^{-9}$ s is of the order of the microsecond. This order of magnitude is in good agreement with our experimental results.\\
 From this first magnetic characterization of our frozen ferrofluid sample, and more precisely from the critical slowing down reported near $T_g$, the transition towards a disordered state observed at low temperature is very likely to be ascribed to a superspin glass transition.

\subsection{ZFCM relaxations}

In order to illustrate what happens to the ZFCM relaxation when the intensity of the probing field $H$ is increased, we have respectively plotted in Figure 2 and Figure 3 the time dependence of the magnetization $M$, normalized by its Field Cooled value $M_{FC}$, and the corresponding relaxation rate $d\,\mathrm{ZFCM}/d\,log(t)$, recorded at $T_m=0.7\ T_g$ for a fixed waiting time $t_w=18000\ s$. The probing field was varied in the range 1 - 8.5 Oe. For a weak field  (1 Oe) a peak of the relaxation rate in Figure 3 is clearly observed at $t \sim t_w$, whereas an increase in the field intensity yields a shift of the peak position towards reduced times $t \sim t_w^{eff}(H)$.
\begin{figure}[!h]
\includegraphics[width=8cm]{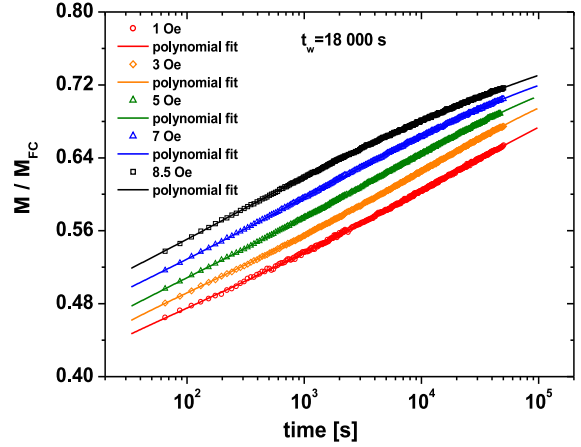}
\caption{\label{fig2} ZFCM relaxations $M(t,t_w=18000s,H)$ normalized by the Field Cooled magnetisation $M_{FC}$. Experiments are performed at $T_{m}=0.7T_{g}$ and here recorded at fixed $t_w=18000\ s$ and increasing probing field $H$. The initial time $t=0$ is defined by the switching on of the magnetic field. A vertical shift (multiplicative coefficient) has been performed for sake of clarity. Full lines are 4$^{th}$ order polynom fits of the data, in order to extract the relaxation rate, $dM/dlog\ t$, presented Fig. 3 }

\end{figure}
\begin{figure}[!h]
\includegraphics[width=8cm]{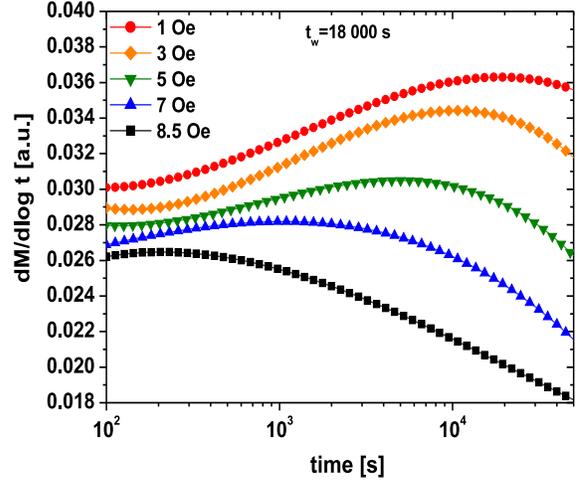}
\caption{\label{fig2} Relaxation rate $dM/dlog\ t$ associated with ZFCM relaxation performed at $T_{m}=0.7T_{g}$ recorded at fixed $t_w=18000\ s$ and increasing probing field $H$. A vertical shift (multiplicative coefficient) has been performed for sake of clarity.}
\end{figure}

Extracting the various $t_w^{eff}(H)$ for different $t_w$ and using the expression of the Zeeman energy $-E_Z/k_BT=ln(t_w^{eff}/t_w)$ we are able to plot on Figure 4 $E_Z(H,N_s(t_w))$ as a function of $H^2$ for the different $t_w$, in $t_w$=3000 - 24000 sec range.  As the temperature equilibration time, after the quench down to 0.7$T_g$, is of the order of 200 seconds, no accurate data is obtained below $t_w=3000$ s. In the probed range of $t_w$, we observe that our data   in superspin glass correspond better to a quadradic dependence of $E_Z$ in $H$ (this behaviour is explicited in the inset of Figure 4). In order to extract the typical number of correlated spins participating in the dynamics at time $t_w$ we performed linear fits of the data to $H^2$ and extracted the slopes. 
The FC susceptibility per superspin, $\chi_{FC}$, of Equation 5 is extracted from the DC value (open circles) of Figure 1, and from the sample concentration and volume. 

\begin{figure}[!h]
\includegraphics[width=8cm]{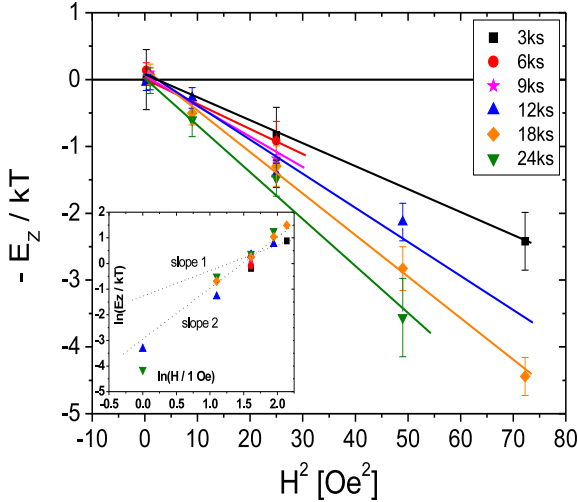}
\caption{\label{fig3} Zeeman energies estimated from ZFCM relaxations vs. field squared for various waiting times in the range [3 ksec - 24ksec]. Inset : logarithmic representation: $ln(E_{Z}/k_{B}T)$ as a function of $ln(H/1Oe)$. Dotted lines with slopes 1 and 2 are added to support the quadratic $H$ field dependence of the Zeeman energy.  }
\end{figure}

The typical number of correlated spins $N_s(t_w)$ found is plotted as a function of $t_w$ in Figure 5. As can be seen, $N_s$ (or equivalently $\xi$) grows with increasing $t_w$. The $N_s$ values found for our frozen ferrofluid, are rather small ($\sim 10^2$), compared to typical values obtained for
spin glasses ($\sim 10^4-10^6$). As discussed above, the dynamical correlation length $\xi$ (normalized to the interparticle distance $\xi_0$) is related to $N_s$ via
\begin{equation}
\frac{\xi}{\xi_0} \sim N_s^{\frac{1}{d-\alpha}}
\label{Eq1} 
\end{equation}
\noindent where $d-\alpha$ stands for the effective dimension of the groups of spins flipping coherently. Whatever the precise value of $d-\alpha$ (between 2 and 3, as indicated by numerical simulations \cite{Berta,Bertb,RiegerBook,Komori:JPSJ99,Marinari:PRL96,Marinari:JPhysA00}), the corresponding values of $\xi$ remain in the order of 5-10 $\xi_0$.

\begin{figure}[!h]
\includegraphics[width=8cm]{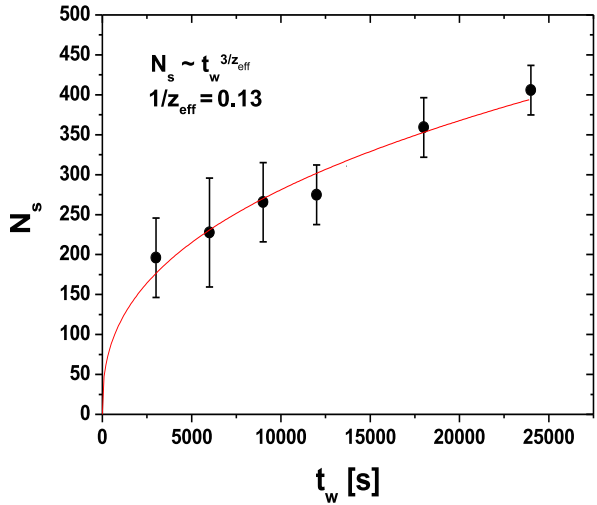}
\caption{\label{fig4} Number of correlated superspins as a function of the waiting time. The solid line is a power law fit of the data.}
\end{figure}

\section{Discussion}

So far, we have been able to apply to the case of a superspin glass the method proposed in \cite{ JohPRL99, DupuisPRL04, Vincent:PRB95} to extract a dynamical correlation length in spin glasses. However, the typical number of correlated superspins is smaller than that obtained in spin glasses \cite{JohPRL99,DupuisPRL04} by 2-3 orders of magnitude. Indeed in \cite{JohPRL99} it was shown that, by simply considering $d-\alpha=3$, the results from several Heisenberg-like atomic spin glasses could be well approximated by 

\begin{equation}
\frac{\xi}{\xi_0}\sim \left(\frac{t_w}{\tau_0} \right)^{\frac{1}{z_{eff}}}
\label{powerlaw}
\end{equation}

\noindent where $\tau_0$ is an individual time and $1/z_{eff}$ is an effective dynamical exponent found to be $\sim\ 0.15T/T_g$. This expression can be seen as an extrapolation below $T_g$ of the dynamic scaling hypothesis, with $z_{eff}(T_g)$ being the usual dynamic exponent $z$. Further experiments on an Ising spin glass \cite{DupuisPRL04} have shown that this simple power law behaviour must be modified to a more general expression, while it can be seen in Fig.5 that our superspin results are already well reproduced by the simple power law Eq.\ref{powerlaw}.

In this expression we see that $\xi$ depends on $t_w$ through the ratio $t_w/\tau_0$. However, if $\tau_0 \sim 10^{-12}\ s$ for individual spins in a spin glass, in superspin glasses, the pertinent individual time given by $\tau_0\sim\bar{\tau_0}e^{\frac{E_a}{k_{B}T}}$ is several orders of magnitude larger than $\bar{\tau_0}$, up to microseconds. Hence, it is clear that we probe a much shorter aging regime of $\xi$ growth in a superspin glass than in an atomic spin glass. This is likely to be the main reason why we observe such small values of $\xi$ in our frozen ferrofluid sample. 

Remarkably, in this regime of smaller numbers of correlated superspins, the growth of the dynamical correlation length follows a power law, with an exponent $1/z_{eff}=0.13\pm0.015$, which is in good agreement with the value $0.15T/T_{g}=0.105$ obtained for Heisenberg-like spin glasses \cite{JohPRL99}. The anisotropy barriers in our nanoparticles are not very large (flipping time $\tau_0$ of the order of hundreds of microseconds in the temperature range of the relaxations), and it is therefore plausible to consider that both superspin experiments and spin glasses simulations explore similar effective aging regimes, $t_w/\tau_0$. Indeed, in Monte Carlo numerical simulations of spin glasses \cite{RiegerBook, Marinari:PRL96, Marinari:JPhysA00, Komori:JPSJ99}, the waiting times are usually of $10^5-10^6$ Monte Carlo Steps (MCS) with an attempt time of 1 MCS. Thus, MC simulations explore a comparable short
aging regime.\\
 However, the \emph{numerical} results for Ising spins  \cite{RiegerBook, Marinari:PRL96, Marinari:JPhysA00, Komori:JPSJ99,Berta} follow a power law behaviour, similar to Eq \ref{powerlaw}, whereas a deviation from such behaviour is found in the Heisenberg case \cite{Bertb}.   On the contrary, in \emph{experiments}, the power law is observed in Heisenberg-like spin glasses, but significant deviations are observed for Ising cases \cite{DupuisPRL04}. The present discrepancy between experiments (in real time) and numerical simulations (in their own distinct time regime) is not yet clearly understood \cite{Bertc}. 

The typical number of correlated superspins, reached in a superspin glass during the experimental timescale, is much smaller than in the case of an atomic spin glass. As noted in \cite{DupuisPRL04,VincentBook,Bouchaud:PRB01} for spin glasses, the number of correlated spins involved in the dynamics must be large enough for such complex phenomena as ``rejuvenation and memory effects'' to occur. This is because these phenomena involve a hierarchy of embedded active length scales. Our results therefore provide a very plausible explanation of the difficulty to observe clear rejuvenation and memory effects in superspin glasses  \cite{Janasson:PRL98, Jonasson:EPJB00, Bouchaud:PRB01, DupuisPRL04}, as already proposed in \cite{Jonsson:PRB05}

\section{Conclusion} 

In this paper, we have reported the results of ZFCM relaxation experiments performed in the superspin glass phase of a concentrated frozen ferrofluid with the anisotropy axis of each nanoparticle randomly orientated. The situation with orientated anisotropy axis will be explored in a forthcoming paper. With the aim of extracting the dynamical correlation length and its age dependence in the superspin glass phase, we used probing fields of increasing intensities and analyzed our data following a method successfully applied to spin glasses. Our results show that the typical number of superspins participating in the slow out-of-equilibrium dynamics (and hence the correlation length) in a superspin glass  follows the same growth law but its value is much smaller than in a spin glass. This difference is likely to be attributed to the much longer individual time, $\tau_{0}$, required for a superspin flip compared to a single spin flip. Consequently, our experiments on a superspin glass allow to probe a much shorter aging regime, $t_w/\tau_0$, of $\xi$ growth than experiments performed on spin glasses. Numerical simulations of spin glasses adapted to superspin glasses (with magnetic dipole-dipole interactions) appear thus as a promising tool to understand strongly interacting nanoparticle systems. They would in addition give new insights in the general problem of the collective behaviour of randomly located dipoles.
\section{Acknowledgments}
The authors deeply thank F. Ladieu and D. l'H\^{o}te, for fruitful discussions, and F. Bert for AC measurements, performed at LPS, University of Paris-Sud. 


\end{document}